\documentclass[prl, twocolumn, superscriptaddress,amsmath,showpacs]{revtex4}
\usepackage{graphicx}
\usepackage{dcolumn}
\usepackage{bm}

\newcommand{\mathcommand}[3][0]{\newcommand{#2}[#1]{\ensuremath{#3}}}
\newcommand{\be}{\begin{equation}}
\newcommand{\ee}{\end{equation}}
\mathcommand[1]{\ket}{\left| #1 \right\rangle}  
\mathcommand[1]{\bra}{\left\langle #1 \right|}  
\mathcommand[2]{\ketbra}{
    \left| #1 \right\rangle\!\!\left\langle #2 \right|} 

\mathcommand[2]{\ts}{#1_{\text{#2}}}
\mathcommand{\te}{\text{e}}
\mathcommand{\thole}{\text{h}}
\mathcommand{\nodag}{{\phantom{\dag}}}

\renewcommand{\vec}[1]{\mathbf{#1}}

\begin{document}
\title{Quantum computers based on electron spins controlled by ultra-fast, off-resonant, single optical pulses}
\par
\author{Susan M. Clark}
\email[Electronic address: ]{sclark4@stanford.edu}
\affiliation{Edward L. Ginzton Laboratory, Stanford University,
Stanford, California 94305-4088, USA}
\author{Kai-Mei C. Fu}
\affiliation{Edward L. Ginzton Laboratory, Stanford University,
Stanford, California 94305-4088, USA}
\author{Thaddeus D. Ladd}
\affiliation{Edward L.
Ginzton Laboratory, Stanford University, Stanford, California
94305-4088, USA}\affiliation{National Institute of Informatics, 2-1-2 Hitotsubashi,
Chiyoda-ku, Tokyo 101-8430, Japan}
\author{Yoshihisa Yamamoto}
\affiliation{Edward L. Ginzton
Laboratory, Stanford University, Stanford, California 94305-4088,
USA}
\affiliation{National Institute of Informatics, 2-1-2 Hitotsubashi,
Chiyoda-ku, Tokyo 101-8430, Japan}

\begin{abstract}
We describe a fast quantum computer based on optically controlled electron spins in charged
quantum dots that are coupled to microcavities. This scheme uses
broad-band optical pulses to rotate electron spins and provide the clock signal to the
system.  Non-local two-qubit gates are performed by phase shifts induced by electron spins on laser pulses
propagating along a shared waveguide.  Numerical simulations of this scheme demonstrate
high-fidelity single-qubit and two-qubit gates with operation times
comparable to the inverse Zeeman frequency.
\end{abstract}

\pacs{03.67.Lx, 
     32.80.Qk, 
     33.35.+r, 
     42.65.Re 
     }

\maketitle

Quantum computers potentially allow improvement in computational speed over existing computers if an architecture is found with a fast clock rate and the ability to be scaled to many qubits and operations~\cite{Nielsen2000}.  Electron spins of charged semiconductor quantum dots are promising candidates for such an architecture because of their potential integration into existing micro-technology.  Most proposals for electron spin quantum computers~\cite{Loss1998, Vrijen2000, Childress2006, Petta2005}, however, restrict logic operations to nearest-neighbors, limiting the computational clock rate.  Optically mediated quantum logic~\cite{Imamoglu1999, Piermarocchi2002, Spiller2006, Yao2005} for two-qubit gates and fast single qubit rotations~\cite{Economou2006, Dutt2006} may improve the overall clock rate of the system.

Several previous works suggest techniques for fast single-qubit rotations of electron spins.  Ground-state coherence generation via ultrafast pulses in molecular, atomic, and quantum dot spectroscopy~\cite{Suter1991, Kis2002, Meshulach1998, Dudovich2002, wu2004, Dutt2006} indicates the ability to control ground state populations and phases.  This control is faster than that of microwave pulses or multiple, adiabatic narrow-band optical pulses.  The application of ultrafast pulses to U(1) control of single quantum-dot qubits has been proposed~\cite{Economou2006}.  Here we describe complete optical SU(2) control of single dots using similar techniques.

There are also proposals for optically-mediated entanglement formation between two non-local qubits.  One type of proposal uses coherently generated single photons~\cite{Yao2005, Cirac1997}, but requires precisely shaped optical pulses.  Recent methods for the entanglement of atomic ensembles via simple optical pulses~\cite{Lukin2000, Duan2000} have led to proposals for optically-mediated two-qubit gates based on small phase shifts of light via single qubits in cavities~\cite{Spiller2006}.  These latter techniques may be easier and faster than the use of coherently generated single photons.  Here, we propose a unique way to combine both fast, SU(2) single-qubit rotations and fast, optically-mediated two-qubit gates on a single semiconductor chip.  These elements may lead toward the fastest potentially-scalable quantum computing scheme of which the authors are aware.

\begin{figure}
\includegraphics[height = .8in]{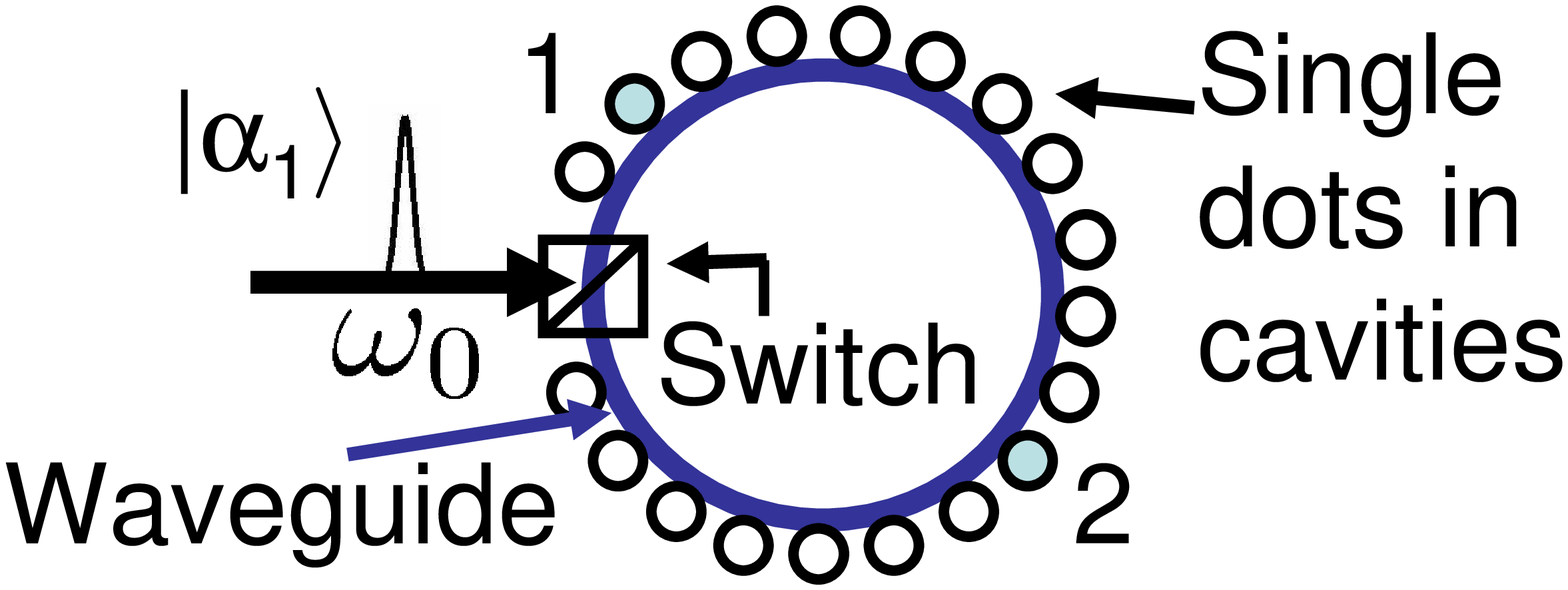}
\caption{Sketch of a loop-qubus quantum computer.  The switch introduces, ejects, and provides displacement operation on the coherent state pulse.
\label{qubus}}
\end{figure}

Figure~\ref{qubus} shows a key component of such an architecture.  It is a square millimeter of a semiconductor chip
patterned with cavities.  Each cavity holds a
single charged quantum dot and is connected to other cavities
through a switched, circular waveguide.  Each quantum dot can be
individually addressed by focused optical pulses incident
perpendicular to the plane of the chip to perform single qubit rotations.  These pulses are part of a pulse
train that serves as the system clock and could be supplied by a semiconductor mode-locked laser~\cite{Sato2001}.
Pulses in the plane of the chip couple distant qubits, forming a ``quantum bus'' or ``qubus,'' which is the foundation of a two qubit gate.

\begin{figure}
\includegraphics[width = 3.375in]{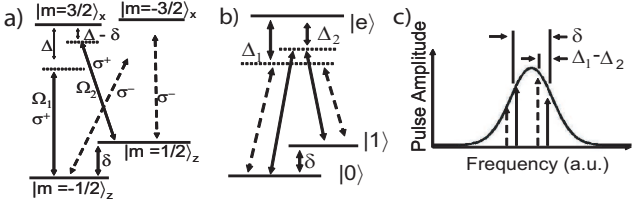}
\caption{{\bfseries a)} Energy level diagram for a charged quantum
dot in an in-plane $B$-field.  Light with $\sigma^{+}$ polarization incident along the growth direction ($x$-axis) couples the
$\ket{m_\te=+1/2}_x=1/\sqrt{2}\left( \ket{m_\te=1/2}_z+\ket{m_\te=-1/2}_z\right)\leftrightarrow\ket{m_\thole=+3/2}_x$
transitions, isolating a three-level system. {\bfseries b)}~Energy
level picture of two pairs of frequencies contained within the
applied pulse that will induce transitions between states $\ket{0}$ and
$\ket{1}$.  {\bfseries c)}~Frequency domain picture of the optical pulse, showing two pairs
of frequency components shown in {\bfseries b)}.}\label{energy}
\end{figure}

We now examine each aspect of this scheme in more detail.  The dots themselves are single-charged, large-area quantum dots (e.g. InGaAs).
Such dots are strong candidates for this architecture because they readily form the three-level
$\Lambda$ system necessary for stimulated Raman transitions
(Fig.~\ref{energy}a) and they have the large oscillator
strengths~\cite{Gammon1996} necessary for fast spin rotations.  The two lower states of this system are the electron Zeeman states and are split by a magnetic field applied along the $z$-direction, which is perpendicular to the growth axis.  The excited state consists of
two electrons in a spin-singlet and a hole. The large
heavy-hole/light-hole splitting allows us to neglect states from the light-hole
excitons and describe the exciton angular-momentum states in the
$x$-basis: $|m_\thole =\pm 3/2\rangle_x$.
If we apply $\sigma^+$-polarized light to the system, the two
electron-spin states, denoted $\ket{0}$ and $\ket{1}$, are
coupled to each other through the single $|m_\thole =+ 3/2\rangle_x$
state, denoted $\ket{\te}$~(Fig.~\ref{energy}a and~b).

\begin{figure}
\includegraphics[width=3in]{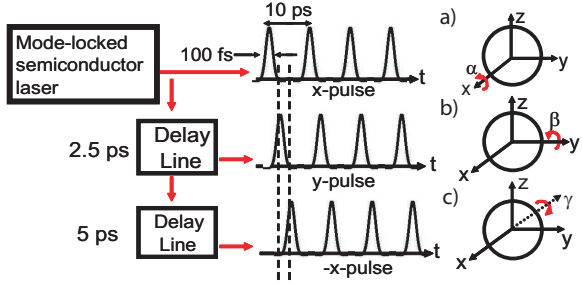}
\caption{Rotations about various axes induced by pulse delays, {\bfseries a)} $x$-pulse train {\bfseries b)} $y$-pulse train {\bfseries c)} $-x$-pulse train.} \label{train}
\end{figure}

Both single qubit and two-qubit gates can be understood from the
rotating-frame Hamiltonian
\be
H = \delta P_1 + \Delta P_\te +
    \sum_{j=0,1}\left[\frac{\Omega_j(t)^\nodag}{2}\sigma_j^+
               +\frac{\Omega_j^\dag(t)}{2}\sigma_j^-\right],
\ee
where $P_j = \ketbra{j}{j}$ is the projection operator for $\ket{j}$ and
$\sigma_j^+ = \ketbra{\te}{j}$ is the raising operator for
$\ket{j}$.  Referring to Fig.~1, $\delta$ is the
ground-state splitting and $\Delta$ is the detuning of the center
frequency of the light pulse from $\omega_0$, the frequency of the
$\ket{0}\rightarrow\ket{\te}$ transition. The meaning of the Rabi frequency
$\Omega_j(t)$ differs in the analyses of single and two-qubit gates. For
single qubit gates, the intense light pulse perpendicular to the cavity
is treated as a classical field and $\Omega_j(t)$ is the product of the dipole matrix element and the time-dependent
electric field amplitude of the the light. For two-qubit
gates, a weak coherent state of light interacts with the quantum
dot in the single-mode cavity and $\Omega_j(t)$ is a
time-dependent Jaynes-Cummings coupling parameter multiplied by the
cavity-photon annihilation operator $a$. There are also incoherent dynamics to be
included in the time evolution of the system, so that the total time
evolution is governed by the master equation
\begin{multline}
\dot\rho = -i[H,\rho]
    -\frac{\Gamma}{2}\biggl(
    P_\te\rho
    +\rho P_e - \sum_{j=0,1}
    \sigma_{j}^-\rho\sigma_j^+\biggr)
    \\-\frac{1}{T_2}\biggl(
    P_1\rho P_0
    +
    P_0\rho P_1
    \biggr).
\label{master}
\end{multline}
Here, $\Gamma$ is the spontaneous emission rate of state $\ket{\te}$
and $T_2$ is the electron spin decoherence rate.

An approximation of the solutions of this equation may
be found by the adiabatic elimination of the excited state, which
is valid when the detuning $\Delta$ is much larger than all other
rates in the system.  The three-level system is then
reduced to the two-level spin system with effective Hamiltonian
$\ts{H}{eff} = G-\ts{\vec{B}}{eff}\cdot\vec{S},$ where $\vec{S}$ is the spin operator of the electron, $G(t)$
generates an irrelevant overall phase, and the effective field is
approximately
\begin{align}
\ts{B}{eff}^z &= -\delta+\frac{\Delta[\Omega_1^\dag\Omega_1^\nodag
                    -\Omega_0^\dag\Omega_0^\nodag]
         +\delta\Omega_0^\dag\Omega_0^\nodag}{4\Delta^2+\Gamma^2}
\label{zterm}
\\
\ts{B}{eff}^x-i\ts{B}{eff}^y &=
    -2\frac{\Delta}{4\Delta^2+\Gamma^2}\Omega_1^\dag\Omega_0^\nodag.
\end{align}
For simplicity, we consider a symmetric $\Lambda$ system with
$\Omega_0=\Omega_1$, in which case $\ts{B}{eff}^y=0$.  Small
deviations from this condition may alter the direction of
\ts{\vec{B}}{eff} during the pulse, but they do not adversely affect
the overall scheme.

For single-spin rotations in which short, intense, highly detuned
pulses are used, the effective field can be much larger than the
applied magnetic field; i.e. the effective Rabi frequency $|\ts{\Omega}{eff}|
=\sqrt{4\Delta^2|\Omega_1|^4/(4\Delta^2+\Gamma^2)^2+\delta^2}\approx
|\Omega_1|^2/2\Delta$ is much faster than the Larmor frequency. The rotation axis is determined by the phase difference between frequency components that are separated by the Zeeman frequency
within the pulse spectrum (Fig.~\ref{energy}b and~c) and thus by the delay time of the pulse
with respect to clock intervals occurring with period $2\pi/\delta$.
To see this, imagine that the spin precesses at Larmor frequency
$\delta$ for a time $\phi/\delta$, at which point an intense,
broadband pulse is applied that rotates it through an angle
$\theta$ about \ts{\vec{B}}{eff}, which is approximately in the $\hat{\vec{x}}$ direction. Then, the spin freely precesses again for a time
$(2\pi-\phi)/\delta$.  This sequence can be written as the unitary
operator
\begin{align}
U =& \exp[i\phi S_z]\exp[-i\theta S_x]\exp[i(2\pi-\phi)S_z]\nonumber\\
  =& -\exp[-i\theta(S_x \cos\phi- S_y \sin\phi)],\label{unitarysingle}
\end{align}
which describes a rotation with an axis determined by $\phi$. Pulses in a pulse train
starting at $t=0$ that arrive at intervals of exactly one Larmor
period cause rotations around the same axis, which we define as the
$x$-axis. Pulses delayed by one fourth or one half of the clock
period will have a phase difference of $\pi/2$ or $\pi$, causing
rotations about the $y$-axis or $-x$-axis, respectively
(Fig.~\ref{train}). This sequence of three pulses can occur in less
than the inverse Zeeman frequency, and thus for a reasonable Zeeman
splitting of 100~GHz, an arbitrary single qubit gate can be
implemented in 10~ps.

To evaluate the importance of terms neglected in our approximate
analysis, we numerically solve Eq.~(\ref{master}) as a three-state
system driven by the classical field $\Omega_1(t)$ using adaptive
Runge-Kutte techniques. We use the realistic quantum dot parameters
${\Gamma}= (200$~ps$)^{-1}$~\cite{Hours2005}, $T_2=
10$~$\mu$s~\cite{Petta2005, Greilich2006}, and $\delta =
100$~GHz.  We assume a Fourier-transform-limited Gaussian pulse
detuned by 10 THz with 100~fs full-width-half-maximum. Using
the definition of fidelity $F = \bra{\psi}\rho \ket{\psi}$, where
$\ket{\psi}$ is the desired quantum state, we find it is possible to
implement both $\pi$- and $\pi/2$-pulses with a fidelity $F>0.999$ (applied
pulse energy densities of 14~$\mu$J/cm$^2$ and 5~$\mu$J/cm$^2$ respectively, which is within the energy output of mode-locked semiconductor lasers followed by optical amplifiers).  The fidelity as a function of Rabi frequency (at the optimal detuning) is shown for both $\pi$- and $\pi/2$ pulses in Fig.~\ref{fid}a.  The general shape of the curve is increasing with Rabi frequency, as larger Rabi frequencies allow for larger detunings and therefore excite less population into the excited state.  There are also some oscillations visible in the curve that are related to Rabi oscillations as the system finds optimum detuning regions. The high fidelity of the single-pulse
Raman rotation is due to the speed of the pulse;
all relaxation and decoherence processes occur at a time scale much
slower than the pulse time.

\newcommand{\kc}{\kappa_{\text{\textsc{c}}}}
\newcommand{\gl}{\Gamma_{\text{\textsc{l}}}}

\begin{figure}
\includegraphics[width = 3 in] {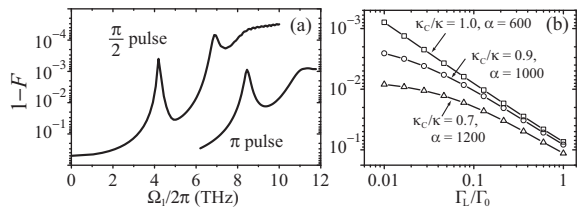}
\caption{{\bfseries a)} Fidelity of single-qubit rotations for $\pi$- and $\pi/2$-pulses vs. Rabi frequency.  {\bfseries b)} Fidelity of two-qubit gates vs. $\gl/\Gamma_0$ for different values of $\kc/\kappa$ and $\alpha$.}\label{fid}
\end{figure}

For two-qubit gates implemented via the ``qubus'' concept, the
single mode cavity is driven by a narrow-band coherent-state pulse.
We assume $\delta \gg \Delta
g^2|\alpha|^2/(4\Delta^2+\Gamma^2)$, where $\alpha$ is the coherent
state amplitude and $g$ is the vacuum Rabi splitting of the
microcavity system. With this assumption, the Hamiltonian is nearly
diagonal and there is negligible population change to the qubit.
According to perturbation theory, the dominant
correction term is the first-order diagonal correction found in the
rightmost term of Eq.~(\ref{zterm}).  This term describes an
effective Hamiltonian of the form $JS^z a^\dag a$, with $J = \delta
g^2/(4\Delta^2+\Gamma^2)$.  This interaction varies the phase of the coherent state field depending
on the spin-state of the quantum dot.  Quantum logic is implemented
by interspersing these optical phase shifts with optical displacements
achieved by mixing the coherent state pulse with a reference pulse
at the optical switch.   Assuming fast, accurate control of the
switching ratio as well as the timing and phase of the reference
pulses, the amplitude of the coherent state may be taken through a
closed path in phase space. The area of this path, and
the resulting geometric phase, depends on the states of the two
qubits interacting with the field, allowing a controlled-phase gate.  Such a gate is
deterministic and does not require detection or feedback.
For details, see Ref.~\onlinecite{Spiller2006}.

The magnitude of the conditional phase shift depends on the detuning
$\Delta$ and the coherent state amplitude $\alpha$.  If $\Delta$ is
too small compared to $g^2/\delta$, there is insufficient
selectivity between the two levels. If $\Delta$ is too large,
the magnitude of $J$ becomes too small compared to decoherence
processes.  Increasing $\alpha$ increases the phase
shift, but if $\alpha$ is too large then decoherence due to spontaneous emission and cavity losses becomes stronger.

To verify the magnitude and fidelity of the
phase-shift operation as a function of $\Delta$ and
$\alpha$, we performed simulations of the
interaction described by Eq.~(\ref{master}). Although the
fully-connected $\Lambda$ system employed in this paper is
different from the asymmetric $\Lambda$ system considered in
Ref.~\onlinecite{Ladd2006}, the effective Hamiltonian $JS^z a^\dag
a$ is the same, and thus many of the qualitative conclusions
apply.  Unlike Ref.~\onlinecite{Ladd2006}, however, quantitative
calculations for the present proposal require a
fully-quantum-mechanical description of the cavity field, because
the previous semi-classical approach fails when pulses are too fast.
For our simulations, we use a basis of displaced Fock states,
$D(\alpha)\ket{n}$, where $D(\alpha)=\exp(\alpha a^\dag+\alpha^*
a).$  If a conditional phase shift $\vartheta$ occurs, the quantum
dynamics may be simulated by a space of approximately
$|\alpha\vartheta|$ of these states ($|\alpha\vartheta|=\sqrt{\pi/2}$ for qubus logic).  We use more states than needed in
the calculation to assure numerical accuracy.  For these
calculations, $\Omega_1(t) = gS(t)a$ where $S(t)$ is the convolution
of the input pulse shape with the filter function of the
cavity~\cite{Ladd2006}. Spontaneous emission in the cavity mode may leak into both the
waveguide, with rate $\kc$, and to lossy modes or absorption, with
rate $\kappa-\kc$. The decay rate $\Gamma$ is now taken as the rate
of spontaneous emission into non-cavity modes, $\gl$, plus emission
into the cavity mode that is lost, so $\Gamma =
\gl+(1-\kc/\kappa)F(\Delta)\Gamma_0$, where $\Gamma_0$ is the decay
rate of the dot in the absence of the cavity (taken to be
(200~ps)$^{-1}$), and $F(\Delta)\ll 1$ is the Purcell factor at high
detuning.  For our simulations, we assume a modest cavity $Q$ of
1000 and a cavity mode-volume of one cubic wavelength inside the
semiconductor;  typical parameters for semiconductor
microcavities~\cite{Englund2005}.

The simulations start with the system in the superposition state
$(\ket{0}+\ket{1})/\sqrt{2}\otimes \ket{\ts{\psi}{i}}$,  where
$\ket{\ts{\psi}{i}}$ is the initial state of the coherent-state optical pulse.  The fidelity is then
calculated as the overlap of the final density matrix with some pure
state $(\ket{0}\ket{\psi_0}+\ket{1}\ket{\psi_1})/\sqrt{2}$, where
$\ket{\psi_j}$ are different optical states.
 We find that Gaussian pulses with root-mean-square width much shorter
than 20~ps cause these optical states to vary significantly from the
desired phase-shifted coherent states.  However, for 20~ps pulses
and with detunings of $\Delta = 4$~THz, we are able to find values
of $\alpha$ large enough to assure coherent states shifted by
$|\alpha\vartheta|=\sqrt{\pi/2}$. We find that the final-state
fidelity depends on the cavity
figures-of-merit $\gl/\Gamma_0$ and $\kc/\kappa$, as shown in
Fig.~\ref{fid}b.  If $\gl/\Gamma_0=0.1$, a value consistent with
existing photonic crystal cavities~\cite{Englund2005}, the fidelity
may reach 99.3\%. The fidelity may also be improved by increasing
the pulse length, increasing the $Q$, or decreasing the mode-volume
of the cavity. Lastly, the fidelity of the final gate also depends on optical loss in the waveguide. The analysis in
Ref.~\onlinecite{Ladd2006} indicates that the percent reduction of
gate fidelity is about equal to the percent amount of loss, and will therefore be a critical parameter to optimize
when designing a fault-tolerant architecture.

The time required for two-qubit gate operations is limited by the
pulse width and the pulse propagation time between the two
qubits.  Nonlocal two-qubit gates will therefore take just a few
periods of the 100~GHz system clock.  To allow gates between
arbitrary qubits, qubits must be switched ``on" and ``off" with
respect to their coupling to the probe pulse field.  In the
schematic of Fig.~\ref{qubus}, it is supposed that each cavity is
far off-resonant from the probe pulse field so that the qubit is
``off" with respect to light-mediated two-qubit gates.  To switch
the interaction on for a particular dot, a powerful, focused,
mid-band light source is introduced only at the cavity of interest
to instantaneously tune it to resonance with the probe pulse (but
still detuned by $\Delta$ from the dot) via the optical Kerr
nonlinearity.

Several features of this scheme favor scalability.  The ability to
achieve two qubit gates between arbitrarily distant qubits is a key
advantage, since schemes relying on nearest-neighbor interactions have more difficulty achieving fault-tolerant
operation~\cite{Svore2005}. Another advantage of our approach is
that the two quantum dots participating in two-qubit gates need not
have the same frequency; several THz inhomogeneity is
tolerated, easing the possibility of large-scale
fabrication. Multiple rings of qubits could be integrated on a
single chip and operated in parallel.  Fast measurement of the
qubits could be accomplished by the same conditional-phase shifts
of bright coherent pulses that are measured via homodyne
detection with ordinary photodetectors.

One technical challenge is presented by the $g$-factor inhomogeneity of quantum
dots.  This inhomogeneity necessitates
the use of spin-echo techniques to synchronize each qubit with the
master clock.  This technical burden may be relieved by using donor-bound excitons instead, as these impurity transitions form the
needed $\Lambda$ transition but show improved $g$
homogeneity~\cite{Karasyuk1994, Fu2005}.

In summary, we have outlined a proposal for performing ultra-fast,
optically controlled quantum gates on electron spins in quantum dots
using stimulated Raman scattering and qubit-controlled phase shifts
with single optical pulses.  For the single-qubit rotations, the
optical pulses have a bandwidth large compared to the splitting of
the two lower $\Lambda$ states; for two-qubit gates the pulses must
have a narrower bandwidth, but may still be as short as 20~ps.  The
timing of the optical pulses is precisely controlled
to provide the system's clock signal and control the qubit rotation
axis.  The clock speed of a single qubit gate in this scheme is
limited only by the lower state splitting.  These methods provide
the basis for an ultra-fast, scalable, solid-state, electron spin
based, all-optical quantum computer.

The authors thank H. Wang, S. E. Harris, K. Nemoto,
W. Munro, and D. Press for helpful discussions.
S. Clark was partially supported by the HP Fellowship Program
through the Center for Integrated Systems. This work was financially
supported by the MURI Center for photonic quantum information
systems (ARO/ARDA Program DAAD19-03-1-0199), JST/SORST program for
the research of quantum information systems for which light is used,
``IT program'' MEXT, University of Tokyo, and the ``Qubus quantum
computer program'' MEXT, NII.


\begin{thebibliography}{28}
\expandafter\ifx\csname natexlab\endcsname\relax\def\natexlab#1{#1}\fi
\expandafter\ifx\csname bibnamefont\endcsname\relax
  \def\bibnamefont#1{#1}\fi
\expandafter\ifx\csname bibfnamefont\endcsname\relax
  \def\bibfnamefont#1{#1}\fi
\expandafter\ifx\csname citenamefont\endcsname\relax
  \def\citenamefont#1{#1}\fi
\expandafter\ifx\csname url\endcsname\relax
  \def\url#1{\texttt{#1}}\fi
\expandafter\ifx\csname urlprefix\endcsname\relax\def\urlprefix{URL }\fi
\providecommand{\bibinfo}[2]{#2}
\providecommand{\eprint}[2][]{\url{#2}}

\bibitem[{\citenamefont{Nielsen and Chuang}(2000)}]{Nielsen2000}
\bibinfo{author}{\bibfnamefont{M.~A.} \bibnamefont{Nielsen}} \bibnamefont{and}
  \bibinfo{author}{\bibfnamefont{I.~L.} \bibnamefont{Chuang}},
  \emph{\bibinfo{title}{Quantum Computation and Quantum Information}}
  (\bibinfo{publisher}{Cambridge University Press}, \bibinfo{address}{New
  York}, \bibinfo{year}{2000}).

\bibitem[{\citenamefont{Loss and DiVincenzo}(1998)}]{Loss1998}
\bibinfo{author}{\bibfnamefont{D.}~\bibnamefont{Loss}} \bibnamefont{and}
  \bibinfo{author}{\bibfnamefont{D.~P.} \bibnamefont{DiVincenzo}},
  \bibinfo{journal}{Phys. Rev. A} \textbf{\bibinfo{volume}{57}},
  \bibinfo{pages}{120} (\bibinfo{year}{1998}).

\bibitem[{\citenamefont{Vrijen et~al.}(2000)}]{Vrijen2000}
\bibinfo{author}{\bibfnamefont{R.}~\bibnamefont{Vrijen}} \bibnamefont{et~al.},
  \bibinfo{journal}{Phys. Rev. A} \textbf{\bibinfo{volume}{62}},
  \bibinfo{pages}{012306} (\bibinfo{year}{2000}).

\bibitem[{\citenamefont{Childress et~al.}(2006)}]{Childress2006}
\bibinfo{author}{\bibfnamefont{L.}~\bibnamefont{Childress}}
  \bibnamefont{et~al.}, \bibinfo{journal}{Phys. Rev. Lett.}
  \textbf{\bibinfo{volume}{96}}, \bibinfo{pages}{070504}
  (\bibinfo{year}{2006}).

\bibitem[{\citenamefont{Petta et~al.}(2005)}]{Petta2005}
\bibinfo{author}{\bibfnamefont{J.~R.} \bibnamefont{Petta}}
  \bibnamefont{et~al.}, \bibinfo{journal}{Science}
  \textbf{\bibinfo{volume}{309}}, \bibinfo{pages}{2180} (\bibinfo{year}{2005}).

\bibitem[{\citenamefont{Imamoglu et~al.}(1999)}]{Imamoglu1999}
\bibinfo{author}{\bibfnamefont{A.}~\bibnamefont{Imamoglu}}
  \bibnamefont{et~al.}, \bibinfo{journal}{Phys. Rev. Lett.}
  \textbf{\bibinfo{volume}{83}}, \bibinfo{pages}{4204} (\bibinfo{year}{1999}).

\bibitem[{\citenamefont{Piermarocchi et~al.}(2002)}]{Piermarocchi2002}
\bibinfo{author}{\bibfnamefont{C.}~\bibnamefont{Piermarocchi}}
  \bibnamefont{et~al.}, \bibinfo{journal}{Phys. Rev. Lett.}
  \textbf{\bibinfo{volume}{89}}, \bibinfo{pages}{167402}
  (\bibinfo{year}{2002}).

\bibitem[{\citenamefont{Spiller et~al.}(2006)}]{Spiller2006}
\bibinfo{author}{\bibfnamefont{T.~P.} \bibnamefont{Spiller}}
  \bibnamefont{et~al.}, \bibinfo{journal}{New J. Phys.}
  \textbf{\bibinfo{volume}{8}}, \bibinfo{pages}{30} (\bibinfo{year}{2006}).

\bibitem[{\citenamefont{Yao et~al.}(2005)\citenamefont{Yao, Liu, and
  Sham}}]{Yao2005}
\bibinfo{author}{\bibfnamefont{W.}~\bibnamefont{Yao}},
  \bibinfo{author}{\bibfnamefont{R.-B.} \bibnamefont{Liu}}, \bibnamefont{and}
  \bibinfo{author}{\bibfnamefont{L.~J.} \bibnamefont{Sham}},
  \bibinfo{journal}{Phys. Rev. Lett.} \textbf{\bibinfo{volume}{95}},
  \bibinfo{pages}{030504} (\bibinfo{year}{2005}).

\bibitem[{\citenamefont{Economou et~al.}(2006)}]{Economou2006}
\bibinfo{author}{\bibfnamefont{S.~E.} \bibnamefont{Economou}}
  \bibnamefont{et~al.}, \bibinfo{journal}{Phys. Rev. B}
  \textbf{\bibinfo{volume}{74}}, \bibinfo{pages}{205415}
  (\bibinfo{year}{2006}).

\bibitem[{\citenamefont{Dutt et~al.}(2006)}]{Dutt2006}
\bibinfo{author}{\bibfnamefont{M.~V.~G.} \bibnamefont{Dutt}}
  \bibnamefont{et~al.}, \bibinfo{journal}{Phys. Rev. B}
  \textbf{\bibinfo{volume}{74}}, \bibinfo{pages}{125306}
  (\bibinfo{year}{2006}).

\bibitem[{\citenamefont{Suter and Mlynek}(1991)}]{Suter1991}
\bibinfo{author}{\bibfnamefont{D.}~\bibnamefont{Suter}} \bibnamefont{and}
  \bibinfo{author}{\bibfnamefont{J.}~\bibnamefont{Mlynek}},
  \bibinfo{journal}{Phys. Rev. A} \textbf{\bibinfo{volume}{43}},
  \bibinfo{pages}{6124} (\bibinfo{year}{1991}).

\bibitem[{\citenamefont{Kis and Renzoni}(2002)}]{Kis2002}
\bibinfo{author}{\bibfnamefont{Z.}~\bibnamefont{Kis}} \bibnamefont{and}
  \bibinfo{author}{\bibfnamefont{F.}~\bibnamefont{Renzoni}},
  \bibinfo{journal}{Phys. Rev. A} \textbf{\bibinfo{volume}{65}},
  \bibinfo{pages}{032318} (\bibinfo{year}{2002}).

\bibitem[{\citenamefont{Meshulach and Silberberg}(1998)}]{Meshulach1998}
\bibinfo{author}{\bibfnamefont{D.}~\bibnamefont{Meshulach}} \bibnamefont{and}
  \bibinfo{author}{\bibfnamefont{Y.}~\bibnamefont{Silberberg}},
  \bibinfo{journal}{Nature} \textbf{\bibinfo{volume}{396}},
  \bibinfo{pages}{239} (\bibinfo{year}{1998}).

\bibitem[{\citenamefont{Dudovich et~al.}(2002)\citenamefont{Dudovich, Oron, and
  Silberberg}}]{Dudovich2002}
\bibinfo{author}{\bibfnamefont{N.}~\bibnamefont{Dudovich}},
  \bibinfo{author}{\bibfnamefont{D.}~\bibnamefont{Oron}}, \bibnamefont{and}
  \bibinfo{author}{\bibfnamefont{Y.}~\bibnamefont{Silberberg}},
  \bibinfo{journal}{Nature} \textbf{\bibinfo{volume}{418}},
  \bibinfo{pages}{512} (\bibinfo{year}{2002}).

\bibitem[{\citenamefont{Wu et~al.}(2004)}]{wu2004}
\bibinfo{author}{\bibfnamefont{Y.}~\bibnamefont{Wu}} \bibnamefont{et~al.},
  \bibinfo{journal}{Physica E} \textbf{\bibinfo{volume}{25}},
  \bibinfo{pages}{242} (\bibinfo{year}{2004}).

\bibitem[{\citenamefont{Cirac et~al.}(1997)}]{Cirac1997}
\bibinfo{author}{\bibfnamefont{J.~I.} \bibnamefont{Cirac}}
  \bibnamefont{et~al.}, \bibinfo{journal}{Phys. Rev. Lett.}
  \textbf{\bibinfo{volume}{78}}, \bibinfo{pages}{3221} (\bibinfo{year}{1997}).

\bibitem[{\citenamefont{Lukin et~al.}(2000)\citenamefont{Lukin, Yelin, and
  Fleischhauer}}]{Lukin2000}
\bibinfo{author}{\bibfnamefont{M.~D.} \bibnamefont{Lukin}},
  \bibinfo{author}{\bibfnamefont{S.~F.} \bibnamefont{Yelin}}, \bibnamefont{and}
  \bibinfo{author}{\bibfnamefont{M.}~\bibnamefont{Fleischhauer}},
  \bibinfo{journal}{Phys. Rev. Lett.} \textbf{\bibinfo{volume}{84}},
  \bibinfo{pages}{4232} (\bibinfo{year}{2000}).

\bibitem[{\citenamefont{Duan et~al.}(2000)}]{Duan2000}
\bibinfo{author}{\bibfnamefont{L.-M.} \bibnamefont{Duan}} \bibnamefont{et~al.},
  \bibinfo{journal}{Phys. Rev. Lett.} \textbf{\bibinfo{volume}{85}},
  \bibinfo{pages}{5643} (\bibinfo{year}{2000}).

\bibitem[{\citenamefont{Sato}(2001)}]{Sato2001}
\bibinfo{author}{\bibfnamefont{K.}~\bibnamefont{Sato}},
  \bibinfo{journal}{Electronic Letters} \textbf{\bibinfo{volume}{37}},
  \bibinfo{pages}{763} (\bibinfo{year}{2001}).

\bibitem[{\citenamefont{Gammon et~al.}(1996)}]{Gammon1996}
\bibinfo{author}{\bibfnamefont{D.}~\bibnamefont{Gammon}} \bibnamefont{et~al.},
  \bibinfo{journal}{Science} \textbf{\bibinfo{volume}{273}},
  \bibinfo{pages}{87} (\bibinfo{year}{1996}).

\bibitem[{\citenamefont{Hours et~al.}(2005)}]{Hours2005}
\bibinfo{author}{\bibfnamefont{J.}~\bibnamefont{Hours}} \bibnamefont{et~al.},
  in \emph{\bibinfo{booktitle}{Physics of Semiconductors: $27^{th}$
  International Conference on the Physics of Semiconductors}}
  (\bibinfo{year}{2005}), p. \bibinfo{pages}{771}.

\bibitem[{\citenamefont{Greilich et~al.}(2006)}]{Greilich2006}
\bibinfo{author}{\bibfnamefont{A.}~\bibnamefont{Greilich}}
  \bibnamefont{et~al.}, \bibinfo{journal}{Science}
  \textbf{\bibinfo{volume}{313}}, \bibinfo{pages}{341} (\bibinfo{year}{2006}).

\bibitem[{\citenamefont{Ladd et~al.}(2006)}]{Ladd2006}
\bibinfo{author}{\bibfnamefont{T.~D.} \bibnamefont{Ladd}} \bibnamefont{et~al.},
  \bibinfo{journal}{New J. Phys.} \textbf{\bibinfo{volume}{8}},
  \bibinfo{pages}{184} (\bibinfo{year}{2006}).

\bibitem[{\citenamefont{Englund et~al.}(2005)}]{Englund2005}
\bibinfo{author}{\bibfnamefont{D.}~\bibnamefont{Englund}} \bibnamefont{et~al.},
  \bibinfo{journal}{Phys. Rev. Lett.} \textbf{\bibinfo{volume}{95}},
  \bibinfo{pages}{013904} (\bibinfo{year}{2005}).

\bibitem[{\citenamefont{Svore et~al.}(2005)\citenamefont{Svore, Terhal, and
  DiVincenzo}}]{Svore2005}
\bibinfo{author}{\bibfnamefont{K.~M.} \bibnamefont{Svore}},
  \bibinfo{author}{\bibfnamefont{B.~M.} \bibnamefont{Terhal}},
  \bibnamefont{and} \bibinfo{author}{\bibfnamefont{D.~P.}
  \bibnamefont{DiVincenzo}}, \bibinfo{journal}{Phys. Rev. A}
  \textbf{\bibinfo{volume}{72}}, \bibinfo{pages}{022317}
  (\bibinfo{year}{2005}).

\bibitem[{\citenamefont{Karasyuk et~al.}(1994)}]{Karasyuk1994}
\bibinfo{author}{\bibfnamefont{V.~A.} \bibnamefont{Karasyuk}}
  \bibnamefont{et~al.}, \bibinfo{journal}{Phys. Rev. B}
  \textbf{\bibinfo{volume}{49}}, \bibinfo{pages}{16381} (\bibinfo{year}{1994}).

\bibitem[{\citenamefont{Fu et~al.}(2005)}]{Fu2005}
\bibinfo{author}{\bibfnamefont{K.-M.~C.} \bibnamefont{Fu}}
  \bibnamefont{et~al.}, \bibinfo{journal}{Phys. Rev. Lett.}
  \textbf{\bibinfo{volume}{95}}, \bibinfo{pages}{187405}
  (\bibinfo{year}{2005}).

\end{thebibliography}
\end{document}